\documentclass[aps,prd,twocolumn,floatfix,nofootinbib,superscriptaddress]{revtex4-1}
\usepackage[colorlinks,linkcolor=blue,anchorcolor=blue,citecolor=blue,urlcolor=blue,breaklinks=true]{hyperref}
\usepackage[libertine]{newtxmath}
\usepackage{graphicx}
\usepackage{slashed}
\usepackage{amsfonts}
\usepackage{amsmath}
\usepackage{blindtext}
\usepackage{microtype}
\usepackage{easyReview}
\usepackage{booktabs}
\usepackage{cleveref}

\newcommand{\vect}[1]{\boldsymbol{#1}}

\graphicspath{{.}{figures/}}

\def\g{\gamma}

\def\hs{\hspace}

\def\no{\nonumber}

\def\ua{\uparrow}
\def\da{\downarrow}

\def\lf{\left}
\def\rg{\right}

\begin{document}

\title{Transverse momentum distributions of valence quark in light and heavy vector mesons.}

\author{Chao Shi}
\email[]{cshi@nuaa.edu.cn}
\affiliation{Department of Nuclear Science and Technology, Nanjing University of Aeronautics and Astronautics, Nanjing 210016, China}

\author{Jicheng Li}
\affiliation{Department of Nuclear Science and Technology, Nanjing University of Aeronautics and Astronautics, Nanjing 210016, China}

\author{Ming Li}
\affiliation{Department of Nuclear Science and Technology, Nanjing University of Aeronautics and Astronautics, Nanjing 210016, China}

\author{Xurong Chen}
\affiliation{Institute of Modern Physics, Chinese Academy of Sciences, Lanzhou 730000, China}
\affiliation{Guangdong Provincial Key Laboratory of Nuclear Science, Institute of Quantum Matter, South China Normal University, Guangzhou 510006, China}

\author{Wenbao Jia}
\affiliation{Department of Nuclear Science and Technology, Nanjing University of Aeronautics and Astronautics, Nanjing 210016, China}

\begin{abstract}

We study the leading-twist time-reversal even transverse momentum dependent parton distribution functions (TMDs) of light and heavy vector mesons, i.e., the $\rho$, $J/\psi$ and $\Upsilon$. We employ the leading Fock-state light front wave functions (LF-LFWFs) of  $\rho$ and $J/\psi$ from our recent study, and supplement with $\Upsilon$'s LF-LFWFs. These LF-LFWFs are extracted from dynamically solved Bethe-Salpeter wave functions. The vector meson TMDs are then studied with the light front overlap representation at leading Fock-state. All the obtained TMDs are non-vanishing and evolve with current quark mass, in particular the tensor polarized TMDs $f_{1LT}$ and $f_{1TT}$ which undergo a sign flip. The $\rho$ TMDs are compared with other model studies and agreement is found, aside from $f_{1LT}$ and $f_{1TT}$. Finally, the collinear PDFs of  vector mesons are studied. The $\rho$'s valence PDFs $f_{1,v}(x)$ and $g_{1L,v}(x)$ are evolved to the scale of 2.4 GeV, with their first three moments compared to lattice QCD prediction. The qualitative behavior of tensor polarized PDF $f_{1LL}(x)$ in $\rho$ at large $x$ is also discussed.
\end{abstract}
\maketitle

%===============================================================================
%===============================================================================
\section{INTRODUCTION\label{intro}}
Multi-dimensional imaging of hadrons has excited a lot of interest for the last decades. The transverse momentum-dependent parton distributions (TMDs), in this connection, provide an important extension to the one-dimensinal parton distribution functions (PDFs) by incorporating the transverse motion of the partons and spin-orbit correlations \cite{Sivers:1989cc,Boer:1997nt,Barone:2001sp, Angeles-Martinez:2015sea}.  The TMDs of pion and nucleon, which are spin-0 and spin-1/2 respectively, have thus received extensive studies from  phenomenological models \cite{Pasquini:2005dk,Bacchetta:2008af,Pasquini:2008ax,Pasquini:2014ppa,Noguera:2015iia, Shi:2018zqd} and lattice QCD \cite{Engelhardt:2015xja,Ebert:2019okf,LatticeParton:2020uhz,Li:2021wvl}. Experimentally they can be studied with the Drell-Yan  or semi-inclusive deep inelastic scattering (SIDIS)  processes \cite{Bacchetta:2006tn,Wang:2017zym, Bacchetta:2017gcc,Scimemi:2017etj,Vladimirov:2019bfa,Bury:2020vhj}.  

Meanwhile, the parton distributions of spin-1 particle  had also been studied in the literature. Starting with the one-dimensional case, a new PDF, i.e., the tensor polarized PDF $b_1^q(x)$ (the superscript $q$ refers to quark), emerges in spin-1 target at leading twist \cite{Hoodbhoy:1988am}. It is interpreted as the difference between unpolarized quark distribution function in $\Lambda=0$ and $|\Lambda|=1$ targets, where the $\Lambda$ is the helicity. The sum rule $\int dx [b_1^q(x)-b_1^{\bar{q}}(x)]=0$ thus holds, as the total valence quark in $\Lambda=0$ and $|\Lambda|=1$ targets should be equal \cite{Efremov:1981vs, Close:1990zw}.  The tensor structure function $b_1(x)$  of deuteron is then measured by HERMES collaboration, and found to be nonzero at low $x (x<0.1)$  \cite{HERMES:2005pon}. In the three-dimensional case, the TMDs and TMD fragmentation functions of spin-one target are introduced \cite{Bacchetta:2000jk}. The Soffer bound is then generalized to the case of spin-1 and positivity bounds on TMDs and TMD FFs are obtained \cite{Bacchetta:2001rb}. While no experimental measurement on spin-1 TMDs is available at present, theoretical study can provide insight into the 3d structure of spin-1 particles in the momentum space. The $\rho$ meson TMDs have thus been studied with NJL model \cite{Ninomiya:2017ggn} and light front models \cite{Kaur:2020emh}, and some photon TMDs are studied with the basis light front quantization (BLFQ) approach \cite{Nair:2022evk}. 

In this work, we study the TMDs of both light and heavy vector mesons, i.e., $\rho$, $J/\psi$ and $\Upsilon$. Among them, the $\rho$ meson is constituted from light quarks, with its mass mostly generated by the dynamical chiral symmetry breaking of QCD at low energy \cite{Maris:1997hd,Maris:1999nt}. The $J/\psi$ and $\Upsilon$, on the other hand, gain their masses mostly from the current quark mass generated by the Higgs mechanism.  Meanwhile, the parton motion within $\rho$ is highly relativistic, while in the $J/\psi$ and $\Upsilon$ it is much slower and  non-relativistic, in particular in the $\Upsilon$. By studying the TMDs of light and heavy vector mesons simultaneously, one can investigate the TMDs in both the relativistic and non-relativistic limit, as well as the DCSB effect in shaping them.

There are generally two approaches to calculate the vector meson TMDs, i.e., the covariant approach and the light front approach. The former calculates the covariant Feynman diagrams, such as in \cite{Ninomiya:2017ggn}, and the later resorts to overlap representations in the light front QCD. Both approaches have their own advantages. For instance, the covariant approach could circumvent the direct calculation of higher Fock-state light front wave functions (LFWFs), which has been a hard task in practice. On the other hand, the light front overlap representation approach gives a direct parton picture in terms of the LFWFs. The parton and parent hadron polarization can be read off explicitly, along with the orbital angular momentum (OAM) transfer among them \cite{Pasquini:2008ax}. Meanwhile, positivity bounds can be conveniently derived in the light front approach, and hence automatically satisfied in light front model studies 
\cite{Bacchetta:1999kz,Bacchetta:2001rb}. In this work, we utilize the light front overlap representation at leading Fock-state approximation and study the vector meson TMDs at leading twist.

The LF-LFWFs of vector meson employed in this study are extracted from their covariant Bethe-Salpeter (BS) wave functions. The basic idea is to project the BS wave functions on to the light front \cite{tHooft:1974pnl,Liu:1992dg,Burkardt:2002uc}. It has been demonstrated with the pseudoscalar mesons \cite{Mezrag:2016hnp,Shi:2018zqd,dePaula:2020qna} and then generalized to the case of vector mesons \cite{Shi:2021taf}. Based on the dynamically solved BS wave functions, which accumulated lots of success in hadron study within the Dyson-Schwinger equations (DSEs) formalism \cite{Roberts:1994dr,Maris:2003vk,Cloet:2013jya, Eichmann:2016yit, Yin:2019bxe,Yin:2021uom,Xu:2021mju}, the parton distribution amplitude \cite{Chang:2013pq, Shi:2014uwa,Shi:2015esa}, generalized parton distributions (GPDs) and TMDs of pseudoscalar mesons \cite{Raya:2021zrz,Shi:2020pqe} are further studied. Exclusive process is also studied, e.g., without introducing any new parameters, the $\rho$ and $J/\psi$ LF-LFWFs are put into the color dipole model study of diffractive vector meson productions in e-p collison, and agreement is found with data from HERA \cite{Shi:2021taf}. In this work, we extend our study to vector meson TMDs. Since the light and heavy vector mesons can be studied consistently in the DSEs, we also include the $\Upsilon$ meson. As we will show, the $\Upsilon$ is well dominated by the leading Fock-states, and therefore provides a benchmark for TMDs in the non-reletivisitic limit, which has not been reported in the literature before.

This paper is organized as follows. In section \ref{sec:LF-LFWF} we introduce the BSE-based LF-LFWFs of vector mesons and supplement with the $\Upsilon$ case. We  then recapitulate the definition of vector meson TMDs and their overlap representation in section \ref{sec:overlap}. The TMDs of heavy and light vector mesons are reported in section \ref{sec:re}, where comparison will be made with other model studies. The collinear PDFs will also be studied. Finally we summarize in section \ref{sec:sum}. 

\section{Vector Meson LF-LFWFs from Bethe-Salpeter wave functions \label{sec:LF-LFWF}}

The extraction of vector meson LF-LFWFs from BS wave functions has been introduced with detail in \cite{Shi:2021taf}. Here we recapitulate the formalism. In light front QCD, the leading Fock-state expansion of a vector meson reads 
\begin{align}\label{eq:LFWF1}
	|M\rangle^\Lambda &= \sum_{\lambda,\lambda'}\int \frac{d^2 \vect{k}_T}{(2\pi)^3}\,\frac{dx}{2\sqrt{x\bar{x}}}\, \frac{\delta_{ij}}{\sqrt{3}} \nonumber  \\
	&\hspace{10mm} \Phi^\Lambda_{\lambda,\lambda'}(x,\vect{k}_T)\, b^\dagger_{f,\lambda,i}(x,\vect{k}_T)\, d_{g,\lambda',j}^\dagger(\bar{x},\bar{\vect{k}}_T)|0\rangle.
\end{align}
The $\Phi^\Lambda_{\lambda,\lambda'}$ is the LF-LFWF of meson with helicity $\Lambda$ and quark (antiquark) with spin $\lambda$ ($ \lambda'$). The $\Lambda=0, \pm 1$ and $\lambda=\ua$ or $\da$, which will be denoted as $\ua=+$ and $\da=-$ for abbreviation in the following. The $i$ and $j$ are color indices. The $\vect{k}_T=(k^x,k^y)$ is the transverse momentum of the quark with flavor $f$, and $\bar{\vect{k}}_T=-\vect{k}_T$ for antiquark with flavor $g$. The longitudinal momentum fraction carried by quark is $x=k^+/P^+$, with $\bar{x}=1-x$ for antiquark. Light-cone four-vector of this paper is defined as $A^{\pm} = \tfrac{1}{\sqrt{2}}(A^0 \pm A^3)$ and $\vect{A}_T=(A^1, A^2)$. 

The vector meson LF-LFWFs can be extracted from their covariant Bethe-Salpeter wave functions with \cite{Shi:2021taf}
\begin{align}\label{eq:chi2phi}
	\Phi^\Lambda_{\lambda,\lambda'}(x,\vect{k}_T)&=-\frac{1}{2\sqrt{3}}\int \frac{dk^- dk^+}{2 \pi} \delta(x P^+-k^+) \nonumber\\
	&\hspace{20mm}\textrm{Tr}\left [\Gamma_{\lambda,\lambda'}\gamma^+ \chi^M(k,P) \cdot \epsilon_\Lambda(P) \right ].
\end{align}
The $\chi^M_\mu(k,P)$ is the BS wave function in the momentum space and the $\epsilon_\Lambda(P)$ is the  meson polarization vector. The $\Gamma_{\pm,\mp}=I\pm \gamma_5$ and $\Gamma_{\pm,\pm}=\mp(\gamma^1\mp i\gamma^2)$ project out corresponding quark-antiquark helicity configurations \footnote{$\Gamma_{\pm,\mp}=I\pm \gamma_5$ refers to $\Gamma_{+,-}=I+\gamma_5$ and $\Gamma_{-,+}=I-\gamma_5$, which is by taking the sign in the same row simultaneously. This notation applies throughout this paper.}. The trace is taken over Dirac, color and flavor spaces. An implicit color factor $\delta_{ij}$ is associated with $\Gamma_{\lambda,\lambda'}$. The flavor index yields unity as we will consider $\rho^+$ ($u\bar{d}$), $J/\psi (c\bar{c})$ and $\Upsilon(b\bar{b})$ in this work.

The $\Phi^{\Lambda}_{\lambda,\lambda'}(x,\vect{k}_T)$'s can be further expressed with six independent scalar amplitudes $\psi(x,\vect{k}_T^2)$'s \cite{Carbonell:1998rj,Ji:2003fw}, i.e.,
\begin{align}
	\hspace{00mm}\Phi_{\pm,\mp}^{0}&=\psi^{0}_{(1)},  \ \ \ \ \ 
	&\Phi_{\pm,\pm}^{0}&=\pm k_T^{(\mp)} \psi^{0}_{(2)}, \label{eq:phi1}\\
	\Phi_{\pm,\pm}^{\pm 1}&=\psi^{1}_{(1)},
	&\Phi_{\pm,\mp}^{\pm 1}&=\pm  k_T^{(\pm)}\psi^{1}_{(2)}, \notag \\
	\Phi_{\mp,\pm}^{\pm 1}&=\pm k_T^{(\pm)}\psi^{1}_{(3)},
	&\Phi_{\mp,\mp}^{\pm 1}&=(k_T^{(\pm)})^2\psi^{1}_{(4)}. \label{eq:phi2}
\end{align}
with $k_T^{(\pm)}=k^x \pm i k^y$. For unflavored vector meson ($J/\psi$ or $\Upsilon$) that has charge parity, further constraints can be found \cite{Shi:2021taf}
\begin{align}\label{eq:psi1}
	\psi_{(i)}^{\Lambda}(x,\vect{k}_T^2)=\psi_{(i)}^{\Lambda}(1-x,\vect{k}_T^2),
\end{align}
except
\begin{align}\label{eq:psi2}
	\psi^{1}_{(2)}(x,\vect{k}_T^2)&=-\psi^{1}_{(3)}(1-x,\vect{k}_T^2).
\end{align}
This reduces the number of independent scalar amplitudes to five. Note that the LF-LFWFs can also be classified by their quark-anti-quark OAM projection along the z-axis, which is denoted by $l_z$ \cite{Ji:2003fw}. The angular momentum conservation in $z$-direction enforces $\Lambda=\lambda+\lambda'+l_z$. Based on Eqs.~(\ref{eq:phi1},\ref{eq:phi2}), the $l_z$ can be $0$, $\pm 1$ and $\pm 2$, which are s-, p- and d-wave LF-LFWFs respectively. 

The $\rho^0$ and $J/\psi$ LF-LFWFs have been presented in \cite{Shi:2021taf}. They are based on BS wave functions under rainbow-ladder truncation. In this case, one does not discriminate between $\rho^0$ and $\rho^{\pm}$ BS wave functions, so the $\rho^0$ LF-LFWFs applies to $\rho^+$ as well. In this work, we supplement with the $\Upsilon$ meson LF-LFWFs, which is obtained in the same way as $J/\psi$. Since $\Upsilon$ is significantly heavier than $J/\psi$, it could help zoom into the non-relativistic limit. The 3-dimensional plots of LF-LFWFs for $\rho^+$, $J/\psi$ and $\Upsilon$ can be found in the Appendix.

\begin{table}[htbp]
	\caption{Calculated masses and decay constants of pseudoscalar and vector mesons based on model setup of \cite{Shi:2021nvg} and \cite{Shi:2021taf}. All units are in GeV. The second and fourth row are based on PDG data \cite{ParticleDataGroup:2020ssz}, with lattice QCD results explicitly indicated by references.}
	\label{tab:para}
\begin{center}
\begin{tabular}{ccccccc}
	\toprule
  & $\pi$ &$\eta_c$ &  $\eta_b$& $\rho$ & $J/\psi$ & $\Upsilon$\\
    \midrule
$m$ & 0.131  & 2.92 & 9.40 & 0.72 & 3.09 & 9.48 \\
$m_{\rm exp}$ & 0.138 & 2.98 & 9.39 & 0.78 & 3.10 & 9.46 \\
$f$ & 0.90 & 0.270 & 0.476 & 0.141 & 0.300 & 0.460 \\
$f_{\rm exp/lQCD}$ & 0.92 & 0.279 \cite{Davies:2010ip} & 0.489 \cite{McNeile:2012qf} & 0.156 & 0.294 & 0.459 \cite{Colquhoun:2014ica} \\
   \bottomrule
\end{tabular}
\end{center}
\end{table}

The interaction model and parameters \cite{Maris:1999nt,Qin:2011xq} are required at the step of solving the quark gap equation and vector meson BS equation. The setup in this work follows exactly that in \cite{Shi:2021taf} for $\rho^+$ and $J/\psi$. The bottom quark were incorporated later in \cite{Shi:2021nvg} for the study of $\eta_b$. Aside from the interaction model, the current quark mass we employed is $m_{u/d}=5$ MeV, $m_c(m_c)=1.33$ GeV and $m_b(m_b)=4.30$ GeV. The calculated meson mass and leptonic decay constants are listed in Table.~\ref{tab:para}. Note that the vector meson leptonic decay constants can be calculated  using the BS wave function $\chi_{\mu}(q;P)$  with \cite{Maris:1999nt}
\begin{align}
f_V m_V=\int^\Lambda \frac{dq^4}{(2\pi)^4} \textrm{Tr}[\gamma \cdot \chi(q;P)], \label{eq:fv1}
\end{align}
and meanwhile reproduced using the LF-LFWF $\phi^{\Lambda=0}_{\pm,\mp}$ with 
\begin{align}
f_V=\sqrt{6}\int_0^1 dx \int^{\Lambda}\frac{d^2 \vect{k}_T}{(2 \pi)^3}  \phi^{\Lambda=0}_{\pm,\mp}(x,\vect{k}_T^2),\label{eq:fv2}
\end{align}
as Eq.~(\ref{eq:fv1}) and Eq.~(\ref{eq:fv2}) are actually equivalent given Eq.~(\ref{eq:chi2phi}).

\section{Definition and overlap representation of vector meson TMDs\label{sec:overlap}}

The TMDs of spin-1 hadrons are defined in connection with the transverse momentum dependent quark correlation function
\begin{align}\label{phi}
\Theta_{\beta \alpha}^{(\Lambda)_{\vect{S}}}(x, \vect{k}_T) &= \int \frac{dz^- \, d^2 \vect{z}_T}{(2\pi)^3} \, 
e^{i\lf(xP^+\,z^- - \,\vect{k}_T \cdot\, \vect{z}_T\rg)}   \no\\
&\hs*{10mm}
\times \ {}_{\vect{S}}\langle P,\Lambda | \overline{\psi}_{\alpha}(0) 
\psi_{\beta}(z^-,\vect{z}_T) | P, \Lambda \rangle_{\vect{S}}.
\end{align}
Here the $P$ is the four-momentum of the hadron with $\vect{P}_T=0$, and $x P^+$ and $\vect{k}_T$ are the longitudinal and transverse momentum carried by the parton. The $\vect{S} = (\vect{S}_T, S_L)$ is the spin quantization axis, and $\Lambda=0,\pm 1$ is the hadron's spin projection on $\vect{S}$ \footnote{Note that the $\Lambda$ in section \ref{sec:LF-LFWF} is defined in the helicity basis, i.e., setting $\vect{S}=(0,0,1)$.}. There is also a gauge link connecting the quark fields, which arises from the gluons. We set it to unity for the study of T-even TMDs as in other studies \cite{Pasquini:2014ppa,Noguera:2015iia,Kaur:2020emh}.

At leading twist, there are nine time-reversal even TMDs entering the parametrization of $\Theta_{\beta \alpha}^{(\Lambda)_{\vect{S}}}(x, \vect{k}_T)$, i.e., \cite{Bacchetta:2000jk,Ninomiya:2017ggn}
\begin{widetext}
\begin{align}
\frac{1}{2} \, {\rm Tr}_D \lf[\g^+ \, \Theta^{(\Lambda)_{\vect{S}}}(x, \vect{k}_T)\rg] &= f_1(x, \vect{k}_T^2) + S_{LL} f_{1LL}(x, \vect{k}_T^2) + \frac{\vect{S}_{LT} \cdot \vect{k}_T}{m_V}\, f_{1LT}(x, \vect{k}_T^2) + \frac{\vect{k}_T \cdot \vect{S}_{TT} \cdot \vect{k}_T}{m_V^2} \, f_{1TT}(x, \vect{k}_T^2),    \\
 \frac{1}{2} \, {\rm Tr}_D \lf[\g^+ \g_5 \, \Theta^{(\Lambda)_{\vect{S}}}(x, \vect{k}_T)\rg]& = \Lambda \lf [S_L\,g_{1L}(x,\vect{k}_T^2) + \frac{\vect{k}_T\cdot \vect{S}_T}{m_V}\, g_{1T}(x,\vect{k}_T^2)  \rg],   \\
\frac{1}{2} \, {\rm Tr}_D \lf[-i \sigma^{+i} \g_5\,\Theta^{(\Lambda)_{\vect{S}}}(x, \vect{k}_T)\rg] &= \Lambda \Biggl[S_T^i\,h_1(x,\vect{k}_T^2) + S_L\,\frac{k_T^i}{m_V}\,h_{1L}^\perp(x,\vect{k}_T^2) + \frac{1}{2\,m_V^2} \lf(2\,k_T^i\,\vect{k}_T \cdot \vect{S}_T - S_T^i\,\vect{k}_T^2\rg) h_{1T}^{\perp}(x,\vect{k}_T^2) \Biggr].
\end{align}
\end{widetext}
with
\begin{align}
S_{LL} &= \lf(3 \Lambda^2 - 2\rg) \left(\tfrac{1}{6} - \tfrac{1}{2}\,S_L^2\right),  \\
S_{LT}^i &=  \lf(3 \Lambda^2 - 2\rg) S_L\, S_T^i, \\
S_{TT}^{ij} &= \lf(3 \Lambda^2 - 2\rg) \lf(S_T^i\,S_T^j - \tfrac{1}{2}\,\vect{S}_T^2 \, \delta^{ij} \rg),
\end{align}
The functions $f$, $g$ and $h$ denote the quark polarization for being unpolarized, longitudinally polarized and transversely polarized respectively. The lower index 1 denotes leading twist, and the $T$ and $L$ refers to the hadron polarization. There are three tensor polarized TMDs $f_{1LL}$, $f_{1LT}$ and $f_{1TT}$ that are specific to spin-one hadron.

The overlap representation of vector meson TMDs in terms of LF-LFWFs have been given by authors in \cite{Kaur:2020emh}, with more details in \cite{Ninomiya:2017ggn,Bacchetta:2001rb}. Introducing the quantity
\begin{align}
A_{\lambda'_{q}\Lambda',\lambda_{q}\Lambda}(x,\vect{k}_T)&=\frac{1}{2(2\pi)^3}\sum_{\lambda_{\bar{q}}}\Phi_{\lambda'_{q},\lambda_{\bar{q}}}^{{\Lambda'}^*}(x,\vect{k}_{T})\Phi_{\lambda_{q},\lambda_{\bar{q}}}^{\Lambda}(x,\vect{k}_T),
\end{align}
the vector meson TMDs overlap representation reads \cite{Kaur:2020emh}
\begin{widetext}
\begin{align}
f_{1}(x,\vect{k}^2_T)&=\frac{1}{3}(A_{+0,+0}+A_{-0,-0}+A_{++,++}+A_{-+,-+}+A_{+-,+-}+A_{--,--}), \label{eq:TMD1}
\\
g_{1L}(x,\vect{k}^2_T)&=\frac{1}{2}(A_{++,++}-A_{-+,-+}-A_{+-,+-}+A_{--,--}), \label{eq:TMD2}
\\
g_{1T}(x,\vect{k}^2_T)&=\frac{m_{V}}{2\sqrt{2}\vect{k}^2_T}\lf(k^{(+)}_T(A_{++,+0}-A_{-+,-0}+A_{+0,+-}-A_{-0,--})+k_T^{(-)}(A_{+0,++}-A_{-0,-+}+A_{+-,+0}-A_{--,-0})\rg), \label{eq:TMD3}\\
h_{1}(x,\vect{k}^2_T)&=\frac{1}{2\sqrt{2}}(A_{++,-0}+A_{-0,++}+A_{+0,--}+A_{--,+0}), \label{eq:TMD4}
\\
h_{1L}(x,\vect{k}^2_T)&=\frac{m_{V}}{2\vect{k}^2_T}\lf(k^{(+)}_T(A_{-+,++}-A_{--,+-})+k^{(-)}_T(A_{++,-+}-A_{+-,--})\rg), \label{eq:TMD5}
\\
h_{1T}(x,\vect{k}^2_T)&=\frac{m_{V}^2}{\sqrt{2}\vect{k}^4_T}\lf((k_T^{(+)})^2(A_{-+,+0}+A_{-0,+-})+(k_T^{(-)})^2(A_{+0,-+}+A_{+-,-0})\rg), \label{eq:TMD6}
\\
f_{1LL}(x,\vect{k}^2_T)&=A_{+0,+0}+A_{-0,-0}-\frac{1}{2}(A_{++,++}+A_{-+,-+}+A_{+-,+-}+A_{--,--}), \label{eq:TMD7}
\\
f_{1LT}(x,\vect{k}^2_T)&=\frac{m_{V}}{2\sqrt{2}\vect{k}^2_T}\lf(k^{(+)}_T(A_{++,+0}+A_{-+,-0}-A_{+0,+-}-A_{-0,--})+k^{(-)}_T(A_{+0,++}+A_{-0,-+}-A_{+-,+0}-A_{--,-0})\rg), \label{eq:TMD8}
\\
f_{1TT}(x,\vect{k}^2_T)&=\frac{m_{V}^2}{2\sqrt{2}\vect{k}^2_T}\lf((k_T^{(+)})^2(A_{++,+-}+A_{-+,--})+(k_T^{(-)})^2(A_{+-,++}+A_{--,-+})\rg), \label{eq:TMD9}
\end{align}
\end{widetext}
To comply with the leading Fock-state truncation in Eqs.(\ref{eq:TMD1}-\ref{eq:TMD9}), we rescale our BSE-based LF-LFWFs so that they normalize to unity, for both $\Lambda=0$ and $\Lambda=\pm 1$ vector mesons respectively i.e., 
\begin{align}
	1&=\sum_{\lambda,\lambda'} \int_0^1 dx \int \frac{d \vect{k}_T^2}{2(2 \pi)^3}  |\Phi^{\Lambda,(\textrm{re})}_{\lambda,\lambda'}(x,\vect{k_T})|^2. \label{eq:N2}
\end{align}
The rescaled LF-LFWFs $\Phi_{\lambda,\lambda'}^{\Lambda=0,(\rm{re})}\! =N_1 \Phi_{\lambda,\lambda'}^{\Lambda=0}$ and $\Phi_{\lambda,\lambda'}^{\Lambda=\pm 1,(\rm{re})}\! =N_2 \Phi_{\lambda,\lambda'}^{\Lambda=\pm 1}$. The $(N_1,N_2)$ are $(1.49,1.72)$, $(1.07,1.09)$ and approximately $(1.0,1.0)$ for $\rho$, $J/\psi$ and $\Upsilon$ respectively. The decreasing $N_1$ and $N_2$ indicate the reduction of potential higher Fock-state contributions from light to heavy mesons. In this way, the vector mesons are approximated as a pair of bounded effective quark and antiquark at certain hadronic scale on the light front. In heavy meson this approximation is good, as the higher Fock-states are suppressed. For light meson as $\rho$, it is less good yet we explore its predictions and compare with other model studies.

\begin{figure}[htbp]
\centering\includegraphics[width=1.0\columnwidth]{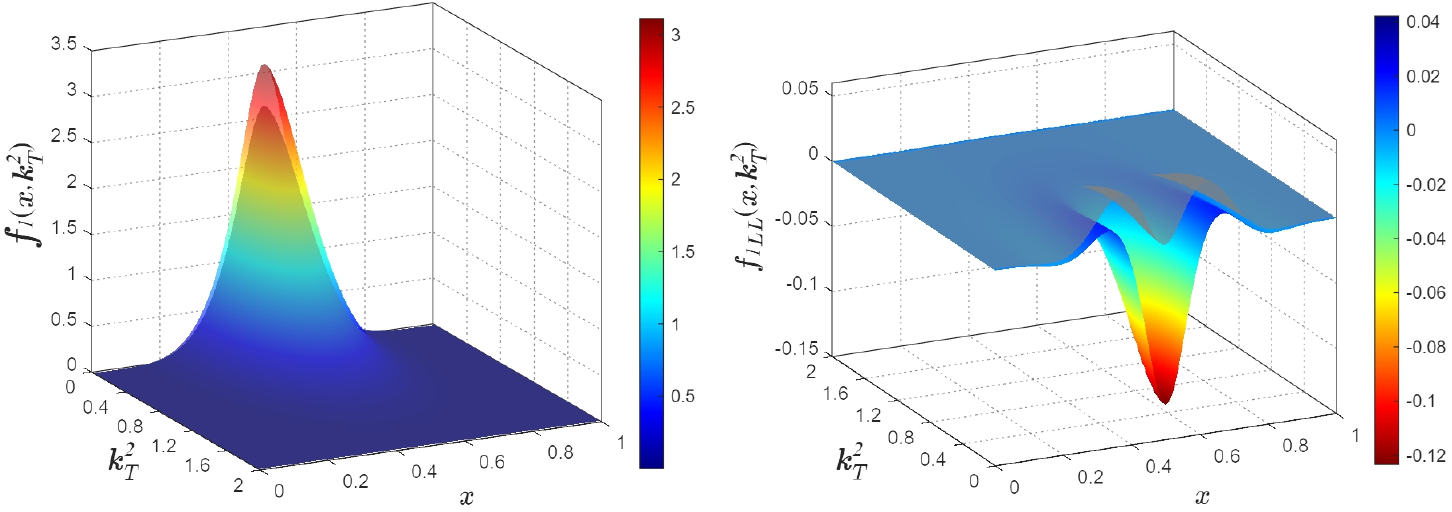} \\
\caption{\looseness=-1 
A demonstration of the $J/\psi$'s $f_{1}(x,\vect{k}_T^2)$ (left column) and $f_{1LL}(x,\vect{k}_T^2)$ (right column) calculated using rescaled (colored surface) and un-rescaled (gray surface) LF-LFWFs. 
}
\label{fig:1}
\end{figure}

Most TMDs are stable in profile under the rescaling procedure, which can be seen from their overlap representation. For instance, in Eqs.(\ref{eq:TMD2},\ref{eq:TMD5},\ref{eq:TMD9}), the $g_{1L}$, $h^\perp_{1L}$ and $f_{1TT}$ only contain overlapping LF-LFWFs of $\Lambda=\pm 1$, so these TMDs change by an overall factor $N_2^2$ after rescaling.  Similarly, the $g_{1T}$, $h_1$, $h^\perp_{1T}$ and $f_{1LT}$, take overlaps between LF-LFWFs of $\Lambda=0$ and $\Lambda=\pm 1$, so they get an overall factor $N_1 N_2$. The $f_1$ and $f_{1LL}$, however, do not have an overall factor. The $f_1$ is the average of the unpolarized TMDs of $\Lambda=0$ and $\Lambda=\pm 1$ meson. Since $N_1$ and $N_2$ doesn't differ much (15\% at most in $\rho$), the profile of $f_1$ doesn't change much either. The $f_{1LL}$, however, is the difference between the unpolarized TMDs of $\Lambda=0$ and $\Lambda=\pm 1$. The cancellation end up being rather sensitive to $N_1/N_2$, and $f_{1LL}$ can change dramatically under the rescaling procedure. This is demonstrated in Fig.~\ref{fig:1} for the case of $J/\psi$. This indicates that $f_{1LL}$ demands careful treatment on higher Fock-state effects, and the rescaling procedure could bring large uncertainties in this respect. We therefore leave $f_{1LL}(x,\vect{k}_T^2)$ out  in this work (for most of the study) and focus on the rest TMDs.

\section{Results\label{sec:re}}

\subsection{TMDs of $J/\psi$ and $\Upsilon$.}

\begin{figure}[htbp]
\centering\includegraphics[width=1.0\columnwidth]{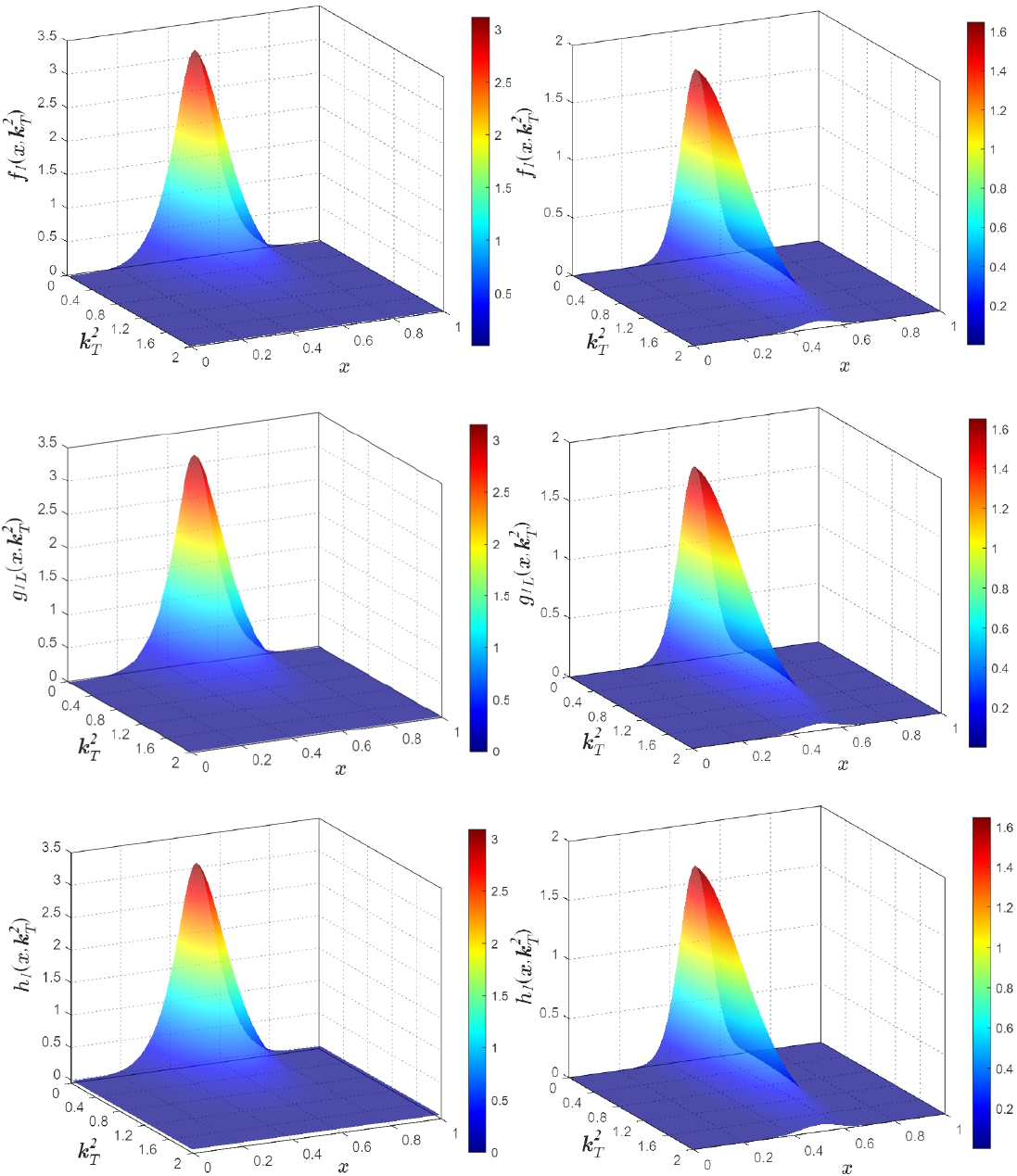} 
\caption{\looseness=-1
3-d plot of the $f_{1}(x,\vect{k}^2_T)$ (top row), $g_{1L}(x,\vect{k}^2_T)$ (middle row) and $h_{1}(x,\vect{k}^2_T)$ (bottom row) for $J/\psi$ (left column) and $\Upsilon$ (right column).
}
\label{fig:2}
\end{figure}

We show in Fig.~\ref{fig:2} the TMDs $f_1(x,\vect{k}_T^2)$, $g_{1L}(x,\vect{k}_T^2)$ and $h_1(x,\vect{k}_T^2)$ of $J/\psi$ (left column) and $\Upsilon$ (right column), which have one-dimensional correspondences, e.g., the collinear PDF $f_1(x)$, $g_{1L}(x)$ and $h_1(x)$ respectively \footnote{We use the same notation for TMD and collinear PDF for convenience. They can be easily distinguished based on the context.}. They describe the momentum distribution of unpolarized, longitudinally polarized and transversely polarized quarks in mesons with same polarization. These TMDs are similar in profile and magnitude within the same meson. They are mostly centered at $x=1/2$ and low $\vect{k}_T^2$ and decrease monotonically, indicating the heavy quark and antiquark tend to have low relative momentum. Meanwhile, $f_1$ is symmetric with respect to $x=1/2$, while the $g_{1L}$ and $h_1$ are slightly asymmetric. The asymmetry originates from the p- and d-wave LF-LFWFs: In Eqs.~(\ref{eq:TMD1},\ref{eq:TMD2},\ref{eq:TMD4}) the overlapping LF-LFWFs are diagonal in $l_z$ for $f$, $g_{1L}$ and $h_1$, so the p- and d-wave contributions to these TMDs can be separated from the s-wave contribution. Given that the p- and d-wave LF-LFWFs are suppressed in heavy mesons, the asymmetry is thus slight. From $J/\psi$ to $\Upsilon$, the TMDs get narrower in $x$ but broader in $\vect{k}_T^2$. Therefore the quark and anti-quark in a heavier meson tend to carry larger transverse momentum but smaller relative longitudinal momentum. 

\begin{figure}[htbp]
\centering\includegraphics[width=1.0\columnwidth]{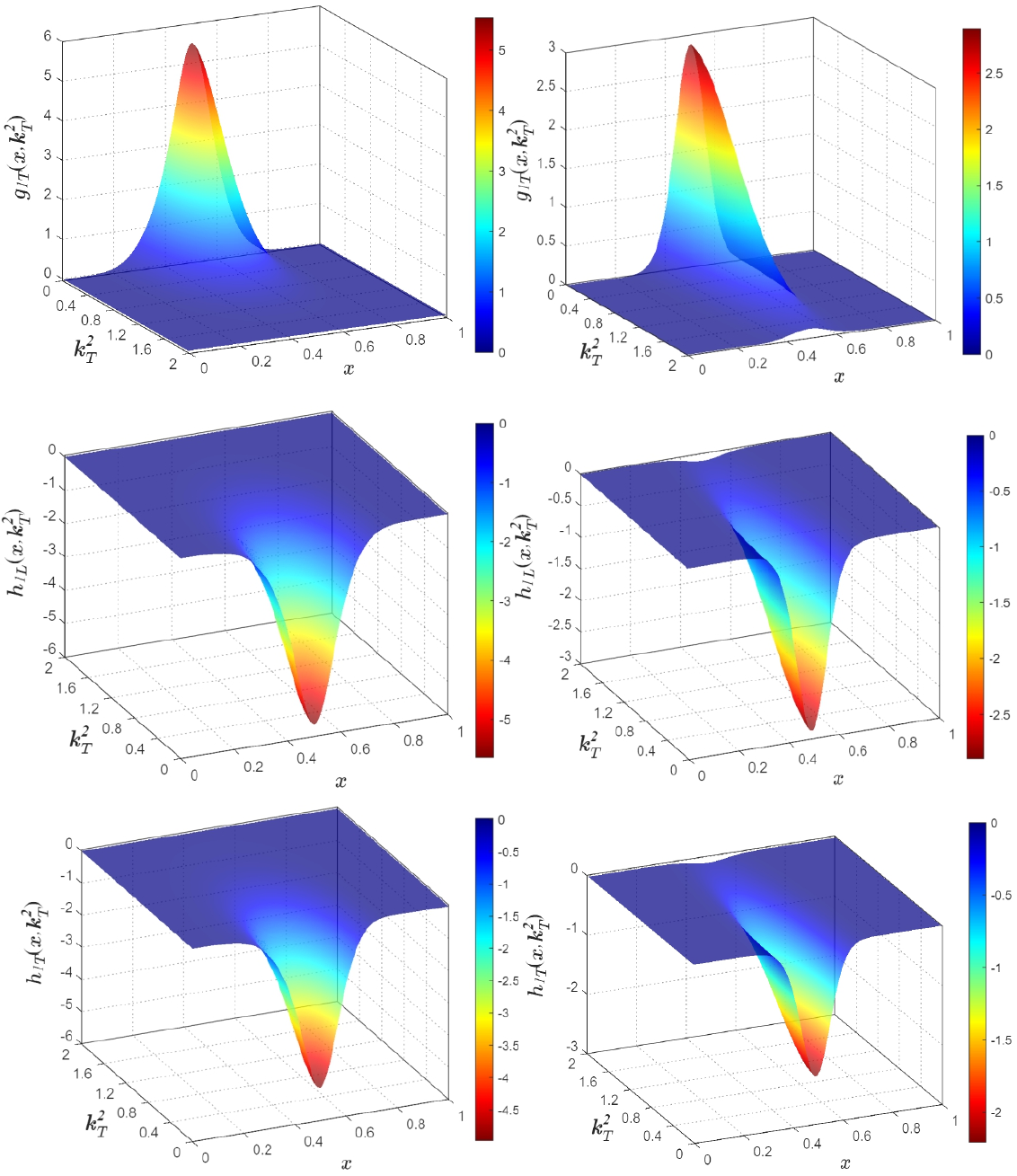} 
\caption{\looseness=-1
3-d plot of the $g_{1T}(x,\vect{k}^2_T)$ (top row), $h_{1L}(x,\vect{k}^2_T)$ (middle row) and $h_{1T}(x,\vect{k}^2_T)$ (bottom row) for $J/\psi$ (left column) and $\Upsilon$ (right column).
}
\label{fig:3}
\end{figure}

%===============================================================================
We then show the worm-gear TMDs $g_{1T}$, $h^\perp_{1L}$ and the pretzelosity TMD $h^\perp_{1T}$ in Fig.~\ref{fig:3}. The $g_{1T}$ describes the momentum distribution of the longitudinally polarized quark in transversely polarized meson, while $h^\perp_{1L}$ and $h^\perp_{1T}$ are for transversely polarized quark in longitudinally and transversely (perpendicular to quark polarization) polarized mesons respectively. These TMDs are similar in magnitude, with \textbf{$h^\perp_{1L}$} and $h^\perp_{1T}$ being negative. They are generally not symmetric in $x\rightarrow 1-x$. From Eqs.~(\ref{eq:TMD3},\ref{eq:TMD5}), one observes that the $g_{1T}$ and $h^\perp_{1L}$ only include overlaps between LF-LFWFs that differ by one unit in OAM, i.e., $|\Delta l_z|=1$, while for $h^\perp_{1T}$ the overlapping LF-LFWFs differ by $|\Delta l_z|=2$. Since our LF-LFWFs are non-vanishing for all possible $l_z$ components, the TMDs are non-vanishing as well. %In the light front holographic model, $\Phi^{\Lambda=0}_{\pm,\mp}$ and $\Phi^{\Lambda=\pm 1}_{\mp,\mp}$ vanish. This excludes a possible nontrivial overlap between $\Phi^{\Lambda=0}$ and $\Phi^{\Lambda=\pm 1}$, hence the $h^\perp_{1T}$ vanishes. 

\begin{figure}[htbp]
\centering\includegraphics[width=1.0\columnwidth]{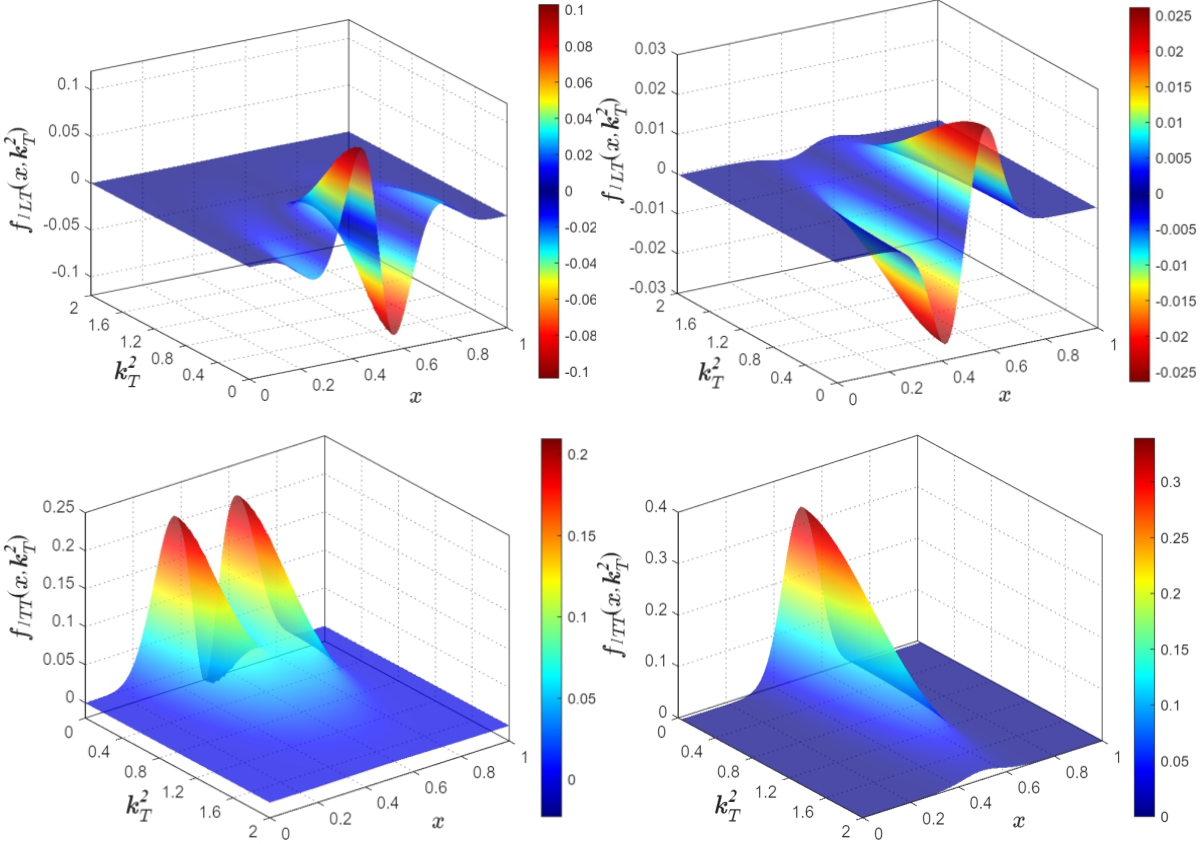} 
\caption{\looseness=-1
3-d plot of the $f_{1LT}(x,\vect{k}^2_T)$ (top row) and $f_{1TT}(x,\vect{k}^2_T)$ (bottom row) for $J/\psi$ (left column) and $\Upsilon$ (right column).
}
\label{fig:4}
\end{figure}

Finally, the tensor polarized TMDs $f_{1LT}$ and $f_{1TT}$ are shown in Fig.~\ref{fig:4}. They are significantly smaller in magnitude as compared to other TMDs. We find $f_{1LT}$ is antisymmetric with respect to $x=1/2$ while $f_{1TT}$ is symmetric. Based on Eqs.(\ref{eq:TMD8},\ref{eq:TMD9}), the $f_{1LT}$ is the overlap between LF-LFWFs of $\Lambda=0$ and $\Lambda=\pm 1$ with one unit of OAM ($\Delta l_z|=1$) transfer while $f_{1TT}$ is the overlap between $\Lambda=+1$ and $\Lambda=-1$ LF-LFWFs with two units of OAM transfer, i.e., $|\Delta l_z|=2$. The $f_{1LT}$ and $f_{1TT}$ are quite different in profile between $J/\psi$ and $\Upsilon$, which was not observed for previously shown TMDs. This indicates that they are sensitive to the current mass of valence quarks in the vector mesons.

\subsection{TMDs of $\rho$.}

\begin{figure*}[htbp]
\centering\includegraphics[width=2.1\columnwidth]{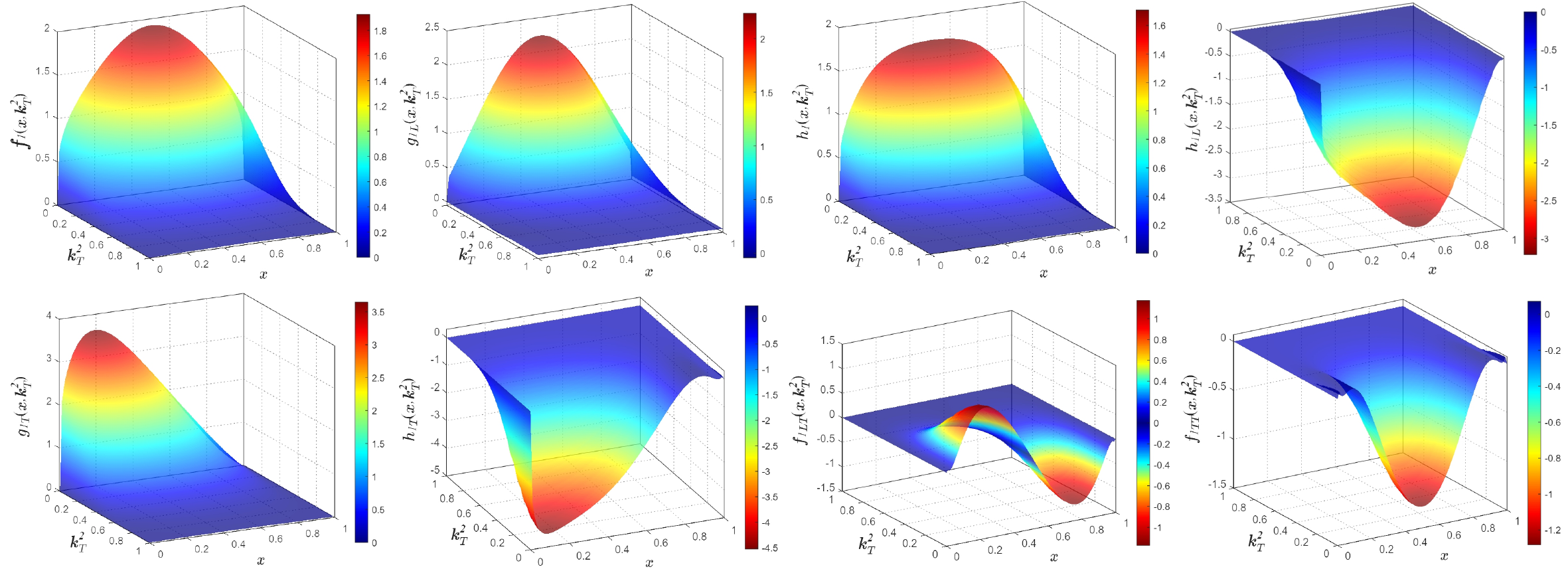} 
\caption{\looseness=-1
The $\rho$ TMDs from the full BSE-based LF-LFWFs.}
\label{fig:5}
\end{figure*}

\begin{figure*}[htbp]
\centering\includegraphics[width=2.1 \columnwidth]{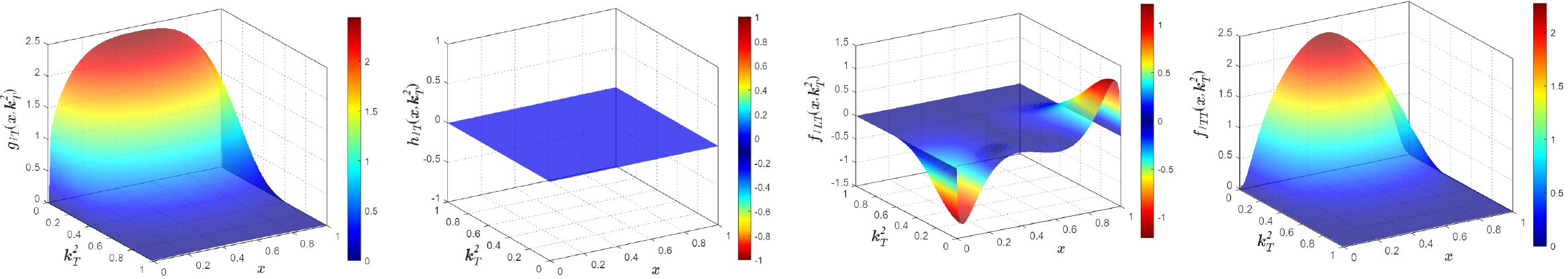}
\caption{\looseness=-1
The $\rho$ TMDs obtained by  setting $\phi^{\Lambda=0}_{|l_z|=1}=\phi^{\Lambda=1}_{|l_z|=2}=0$ in the full BSE-based LF-LFWFs.}
\label{fig:6}
\end{figure*}

We show the calculated $\rho$ TMDs in Fig.~\ref{fig:5}. As compared to the TMDs of heavy vector mesons, they are significantly broader in $x$ and narrower in $\vect{k}_T^2$, following the trend from $\Upsilon$ to $J/\psi$. We remind such effect is also observed in the TMDs of light and heavy pseudoscalar mesons \cite{Shi:2021nvg}. On the other hand, the $\rho$ meson is a highly relativistic system, so the p- and d-wave LF-LFWFs are more pronounced. They bring prominent effect, as the $g_{1L}$, $h_{1}$, $g_{1T}$, $h_{1L}$ and $h_{1T}$ become more asymmetric \footnote{The  $g_{1L}$ and $h_{1}$ are asymmetric at $\vect{k}_T^2 \ne 0$, which would be more obvious by plotting $|\vect{k}_T|g_{1L}$ and $|\vect{k}_T|h_{1}$ instead.}.  Meanwhile, the $f_{1LT}$ and $f_{1TT}$ are strongly enhanced as compared to those in $J/\psi$ and $\Upsilon$. Note that  $f_{1LT}$ and $f_{1TT}$ undergo a flip in sign from $\rho$ to $\Upsilon$.   

It is interesting to compare our $\rho$ TMDs with those from the NJL model \cite{Ninomiya:2017ggn}, the light front (LF) holographic model and the light front (LF) quark model \cite{Kaur:2020emh}. In particular, it is found that the LF holographic  model agree well with NJL model on the profile of all the $\rho$ TMDs, i.e., the TMDs share exactly same behavior such as being vanishing or non-vanishing, positive or negative, as well as the way they are skewed \cite{Kaur:2020emh}. Here we want to point out that, such nice agreement could be due to the fact that the two models actually have LF-LFWFs with the same non-vanishing spin configurations. To see that, we first recapitulate the light front holographic LFWFs of $\rho$, which reads \cite{Forshaw:2012im,Kaur:2020emh}
\begin{align}
\Phi^{\Lambda=0}_{\lambda,\lambda'}(x,\vect{k}_T)&=N_L \delta_{\lambda,-\lambda'}(m_\rho^2 x(1-x)+m_q^2+\vect{k}_T^2)\frac{\psi(x,\vect{k}_T^2)}{x(1-x)}, \label{eq:Phi1}\\
\Phi^{\Lambda=\pm 1}_{\lambda,\lambda'}(x,\vect{k}_T)&=N_T [|\vect{k}_T|e^{\pm i \theta_{\vect{k}_T}}(\pm x\delta_{\lambda \pm,\lambda'\mp}\mp (1-x)\delta_{\lambda \mp,\lambda' \pm}) \nonumber \\
&+m_q \delta_{\lambda \pm,\lambda' \pm}]\frac{\psi(x,\vect{k}_T^2)}{x(1-x)}. \label{eq:Phi2}
\end{align}
Here $\delta_{ab,cd}=\delta_{a,b}\delta_{c,d}$, with $\delta_{a,b}$ the Kronecker delta. The spin configurations are generated by the bare vertex $\gamma_\mu$ with 
\begin{align}\label{eq:spin}
\frac{\bar{u}_{\lambda}(k^+,\vect{k}_T)}{\sqrt{x}}\epsilon_{\Lambda} \cdot \gamma \frac{v_{\lambda'}(k'^+,\vect{k}_T')}{\sqrt{1-x}}.
\end{align}
The $k$ and $k'$ denotes the 4-momenta of the quark and antiquark respectively. They satisfy $k^+=x P^+$, $k'^+=(1-x)P^+$ and $\vect{k}_T=-\vect{k}_T$, where $P$ is the meson four momentum. 
Comparing with Eqs.~(\ref{eq:phi1},\ref{eq:phi2}), one finds that $\Phi^{\Lambda=0}_{\pm,\pm}$  and $\Phi^{\Lambda=\pm 1}_{\mp,\mp}$ vanish in Eqs.(\ref{eq:Phi1},\ref{eq:Phi2}). Note that $\Phi^{\Lambda=0}_{\pm,\pm}$  and $\Phi^{\Lambda=\pm 1}_{\mp,\mp}$ correspond to $|l_z|=1$ and $|l_z|=2$ respectively, so we will denote them as $\Phi^{\Lambda=0}_{|l_z|=1}$  and $\Phi^{\Lambda=\pm 1}_{|l_z|=2}$ in the following. On the other hand, the NJL model calculation takes the covariant formalism rather than the light front overlap formalism. But we can follow formula (\ref{eq:spin}), or equivalently Eq.~(\ref{eq:chi2phi}), and project out the LF-LFWFs from $\rho$'s Bethe-Salpeter amplitude in the NJL model. We remind that the Dirac structure of $\rho$'s BS amplitude contains only the $\gamma_\mu$ term (see Eq.~(47) in \cite{Ninomiya:2017ggn}), so it generates exactly the same spin configurations as LF holographic model does. 

To make an analogous comparison with the LF holographic  model, we set $\Phi^{\Lambda=0}_{|l_z|=1}$ and $\Phi^{\Lambda=\pm 1}_{|l_z|=2}$ of the BSE-based LF-LFWFs to zero, and re-calculate all the TMDs.  While some TMDs do not change much, significant deviations are found in others, which we have picked out and plotted in Fig.~\ref{fig:6}. One can see that from the second row of Fig.~\ref{fig:5} to Fig.~\ref{fig:6}, the $g_{1T}$ gets from asymmetric in $x$ to symmetric, and the $h^\perp_{1T}$ gets from nonvanishing to vanishing. Moreover, the $f_{1LT}$ and $f_{1TT}$ both undergo a sign flip. The LF-LFWFs with higher OAM thus have sizable effect in determining these TMDs. It is worth noting that the TMDs in Fig.~\ref{fig:6} plus the first row of Fig.~\ref{fig:5} agree very well with LF holographic  model or NJL model regarding their profiles. 

Finally, we remark that the LF quark model incorporates nonvanishing $\Phi^{\Lambda=0}_{|l_z|=1}$ and $\Phi^{\Lambda=\pm 1}_{|l_z|=2}$  \cite{Yu:2007hp,Qian:2008px}, and have yield $g_{1T}$ and $h^\perp_{1T}$ that are similar to ours in Fig.~\ref{fig:5} \cite{Kaur:2020emh}. However, the tensor polarized TMDs $f_{1LT}$ and $f_{1TT}$ vanish in the LF quark model, which is different from our result. So at this stage, new possibilities regarding the profile of $f_{1LT}$ and $f_{1TT}$ are presented in Fig.~\ref{fig:5}.

\subsection{Integrated TMDs}
To quantify the transverse momentum dependence in TMDs, we calculate the mean transverse momentum of certain TMDs, defined through
\begin{align}
	\langle \vect{k}_{T}\rangle&=\frac{\int dx\,\ d^2\vect{k}_T|\vect{k}_T|{\cal F}(x,\vect{k}^2_T)}{\int dx\,\ d^2\vect{k}_T {\cal F} (x,\vect{k}^2_T)}, \label{eq:kt}
\end{align}
where the ${\cal F}$ stands for various TMDs. In Table.~\ref{tab:kt} we list the $\langle \vect{k}_{T}\rangle$ of concerned TMDs, and compare with other model studies. Our result is listed in the last three columns. We notice that our results on $\rho$ are generally larger than other model predictions, but close to NJL model for most TMDs.

\begin{table}[htbp]
	\caption{The $\vect{k}_{T}$-moment of TMDs defined in Eq.~(\ref{eq:kt}). Lines in the blank indicate the corresponding TMDs are vanishing. All units are given in GeV. The first three columns are taken from \cite{Kaur:2020emh,Ninomiya:2017ggn}}
	\label{tab:kt}
	\begin{center}
\begin{tabular}{ccccccc}
	\toprule
 & $\langle \vect{k}_{T}\rangle^\rho_{\rm NJL}$ & $\langle \vect{k}_{T}\rangle^\rho_{\rm LFHM}$ & $\langle \vect{k}_{T}\rangle^\rho_{\rm LFQM}$ & $\langle \vect{k}_{T}\rangle^\rho_{\rm BSE}$ & $\langle \vect{k}_{T}\rangle^{J/\psi}_{\rm BSE}$ & $\langle \vect{k}_{T}\rangle^\Upsilon_{\rm BSE}$\\
    \midrule
$f_{1}$ & 0.32 & 0.238 & 0.328 & 0.399 & 0.623 & 1.020\\
$g_{1L}$ & 0.08 & 0.204 & 0.269 & 0.318 & 0.589 & 1.003\\
$g_{1T}$ & 0.34 & 0.229 & 0.269 & 0.358 & 0.615 & 1.020\\
$h_{1}$ & 0.34 & 0.229 & 0.307 & 0.367 & 0.608 & 1.012\\
$h_{1L}$ & 0.33 & 0.204 & 0.269 &0.368 & 0.608 & 1.017\\
$h_{1T}$ & — & — & 0.237 & 0.365 & 0.602 & 1.017\\
$f_{1TT}$ & 0.32 & 0.211 & — & 0.338 & 0.764& 1.063\\
   \bottomrule
   \end{tabular}
\end{center}
\end{table}

The collinear PDFs of vector mesons can be obtained by integrating over the $\vect{k}_T$ in TMDs, i.e., 
\begin{align}
{\cal F}(x)&=\int d\vect{k}_T^2 {\cal F}(x,\vect{k}_T^2),
\end{align}
with  ${\cal F}=f_1, g_{1L}, h_1$ and $f_{1LL}$. The $f_1(x)$ has the probabilistic interpretation of finding an unpolarized quark in an unpolarized meson. The helicity PDF $g_{1L}$ is the number density of quarks with helicity $1$ over quarks with helicity $-1$ in a meson with helicity $1$, and the transversity PDF $h_1$ is the analogue when both quark and meson are transversely polarized along the same axis. The $f_{1LL}$ characterizes the difference of unpolarized quark distribution between $\Lambda=0$ and $\Lambda=\pm 1$ states. We note that the heavy meson PDFs have been studied by the BLFQ approach and light front models \cite{Li:2017mlw,Lan:2019img,Li:2021cwv}.

\begin{figure}[htbp]
\centering\includegraphics[width=0.95\columnwidth]{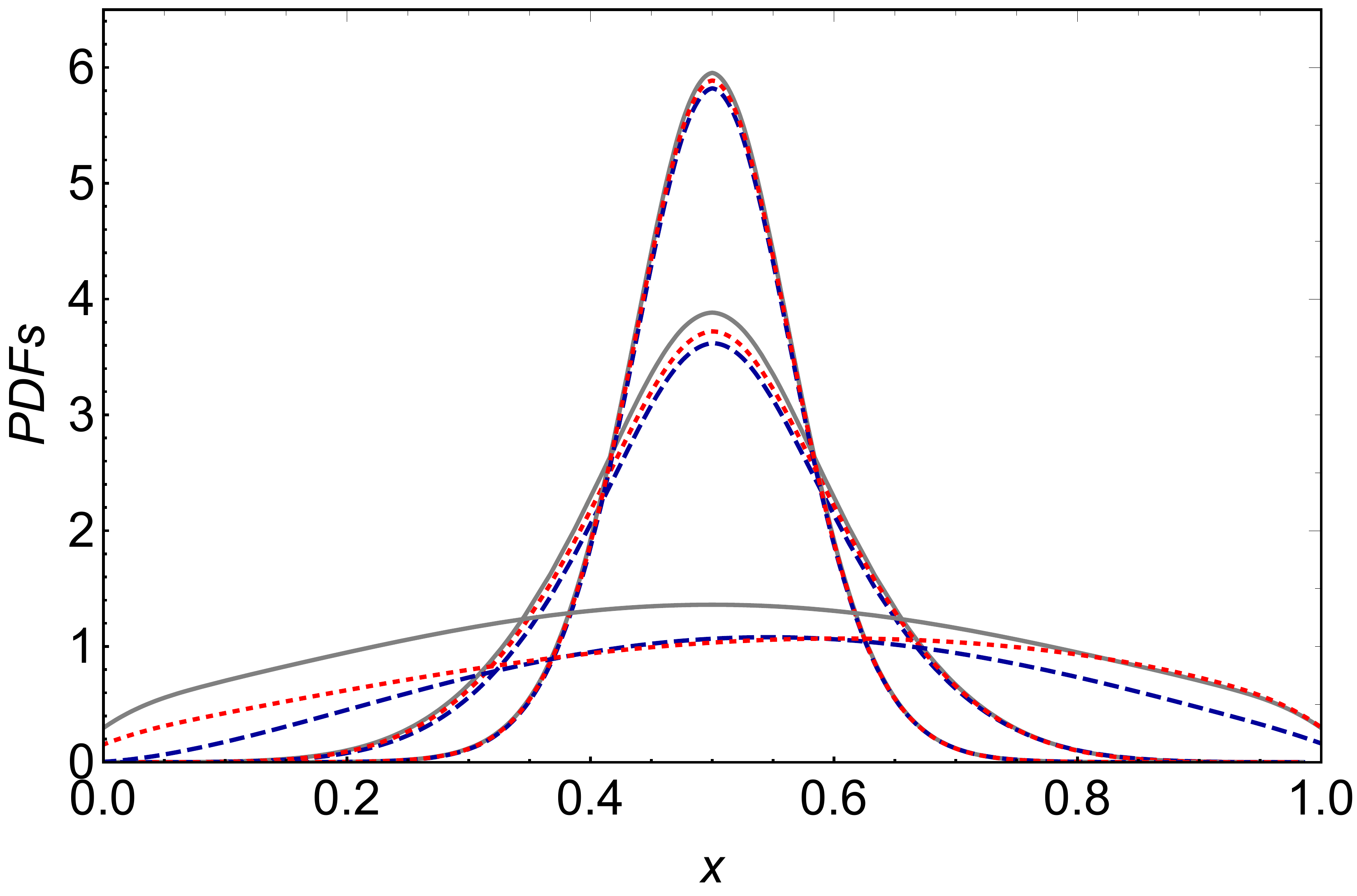} 
\caption{\looseness=-1
The $f_{1}(x)$ (gray solid), $g_{1L}(x)$ (red dotted) and $h_{1}(x)$ (blue dashed) of vector mesons at hadron scale. At $x=0.5$, from top to bottom, the three sets of curves correspond to $\Upsilon$, $J/\psi$ and $\rho$ respectively.
}
\label{fig:7}
\end{figure}

We plot the PDFs $f_1$, $g_{1L}$ and $h_{1}$ of $\rho$, $J/\psi$ and $\Upsilon$ in Fig.~\ref{fig:7}.  The PDFs of heavy mesons are generally narrow in $x$ and centered around $x=1/2$. Meanwhile, PDFs of the same heavy meson are quite close to each other. Looking into the quark spin sum $\langle x^0 \rangle_{g_{1L}}=\int dx g_{1L}(x)$ and tensor charge $\langle x^0 \rangle_{h_1}=\int dx h_{1}(x)$, we find $\langle x^0 \rangle^{J/\psi}_{g_{1L}}=0.92$, $\langle x^0 \rangle^{J/\psi}_{h_1}=0.96$ and $\langle x^0 \rangle^{\Upsilon}_{g_{1L}}=0.98$, $\langle x^0 \rangle^{\Upsilon}_{h}=0.99$. They are less than unity due to non-zero OAM of quarks, and closer to unity in $\Upsilon$ than in $J/\psi$, as the relativistic effect reduces. On the other hand, the PDFs of $\rho$ are much broader and the deviation between the PDFs are more significant. This indicates the quark and anti-quark in a highly relativistic system as $\rho$ are no longer constrained to carry small relative longitudinal momentum as in heavy mesons. Moreover, nonzero OAM configurations become significant as we find $\langle x^0 \rangle^{\rho}_{g_{1L}}=0.67$ and $\langle x^0 \rangle^{\rho}_{h_1}=0.79$. 

\begin{figure}[htbp]
\centering\includegraphics[width=1.0\columnwidth]{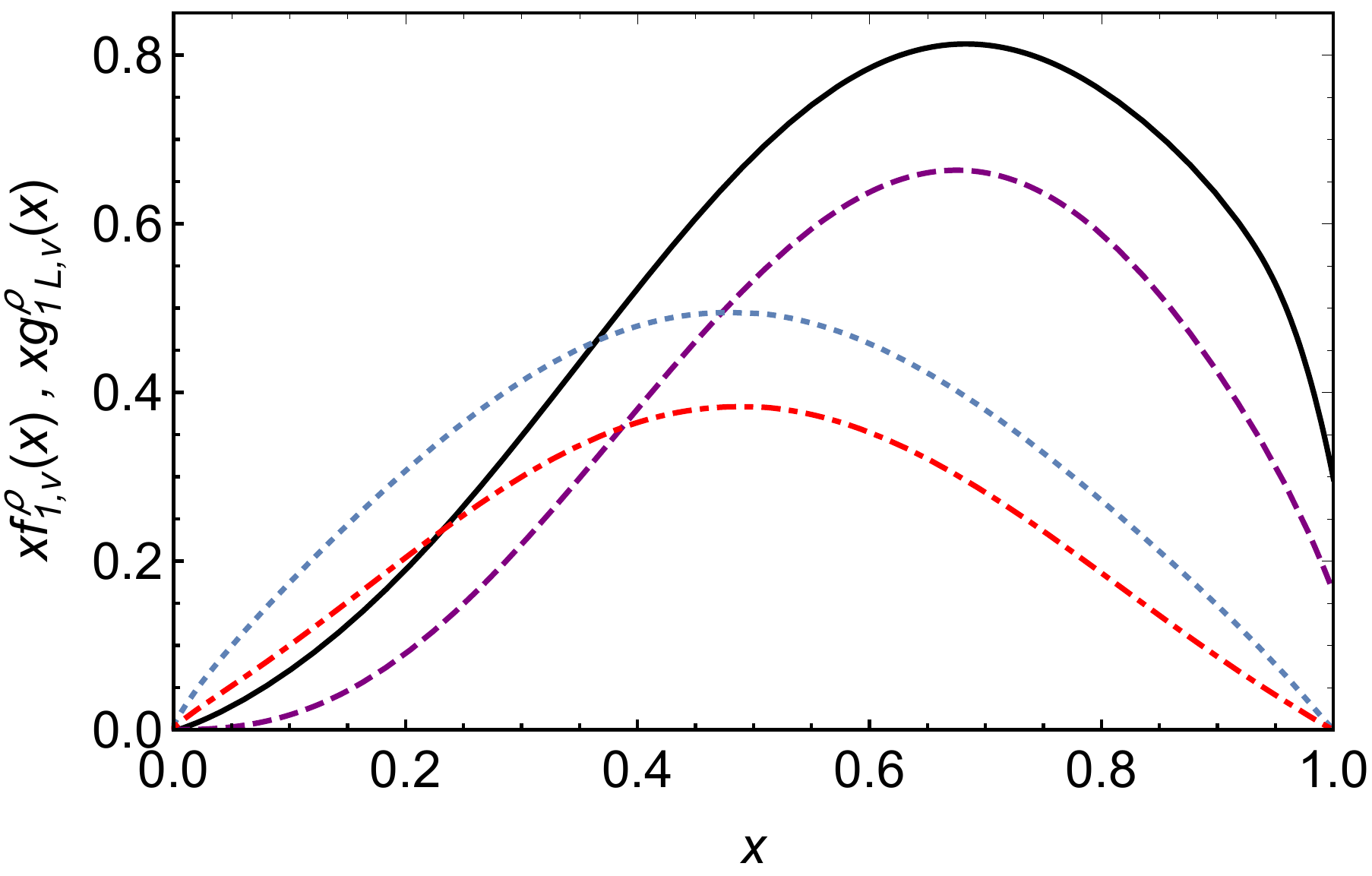} 
\caption{\looseness=-1
The $x f_{1,v}^{\rho}(x)$ at hadronic scale (black solid) and evolved scale of 2.4 GeV (blue dotted), and $x g_{1L,v}^{\rho}(x)$ at hadronic scale (purple dashed) and evolved scale of 2.4 GeV (red dot-dashed).
}
\label{fig:8}
\end{figure}

While the $\rho$ PDFs are unavailable by experiment, lattice QCD has made predictions on their moments \cite{Best:1997qp, Loffler:2021afv}. In \cite{Best:1997qp}, the first three moments of $\rho$'s valence (nonsinglet) unpolarized distribution $a_n=\langle x^{n-1} \rangle_{f_{1,v}}$ and helicity distribution $r_n=\langle x^{n-1} \rangle_{g_{1L,v}}$ at a renormalization scale of $\mu \approx 2.4$ GeV are given  
\begin{align}
a_2&=0.334(21),  a_3=0.174(47), \  a_4=0.066(39) \label{eq:rhoa}\\
r_1&=0.57(32), \ \  r_2=0.212(17),  \ r_3=0.077(34)\label{eq:rhor}
\end{align}
Note in \cite{Best:1997qp}, there are two sets of $r_n$ values extracted from two different operators which should equal in the continuum limit. Here we take their intersection. To compare with the lattice prediction directly, we evolve our PDFs to the scale of $\mu_2=2.4$ GeV using the NLO DGLAP evolution with the help from QCDNUM package \cite{Botje:2010ay}. The strong coupling constant is set to the optimal value in NLO global PDF analysis $\alpha_s(1 \textrm{GeV})=0.491$ \cite{Martin:2009iq} and the variable flavor number scheme (VFNS) is taken. However, the initial scale of our $\rho$ PDFs model is unknown. Here we choose it to be $\mu_0=670$ MeV\footnote{The corresponding expansion parameter of perturbative NLO DGLAP evolution is $\frac{\alpha_s(\mu_0)}{2\pi}=0.126$.}. In this case, the valence $\rho$ PDFs at scales $\mu_0$ and $\mu_2$ are shown in Fig.~\ref{fig:8}, with the later yield 
\begin{align}
a_2&=0.316, \  a_3=0.155, \  a_4=0.091 \\
r_1&=0.66, \ \ \  r_2=0.227,  \ r_3=0.111.
\end{align}
at the scale of 2.4 GeV. They agree with lattice predictions in Eqs.(\ref{eq:rhoa},\ref{eq:rhor}) within uncertainties. 
\begin{figure}[htbp]
\centering\includegraphics[width=1.0\columnwidth]{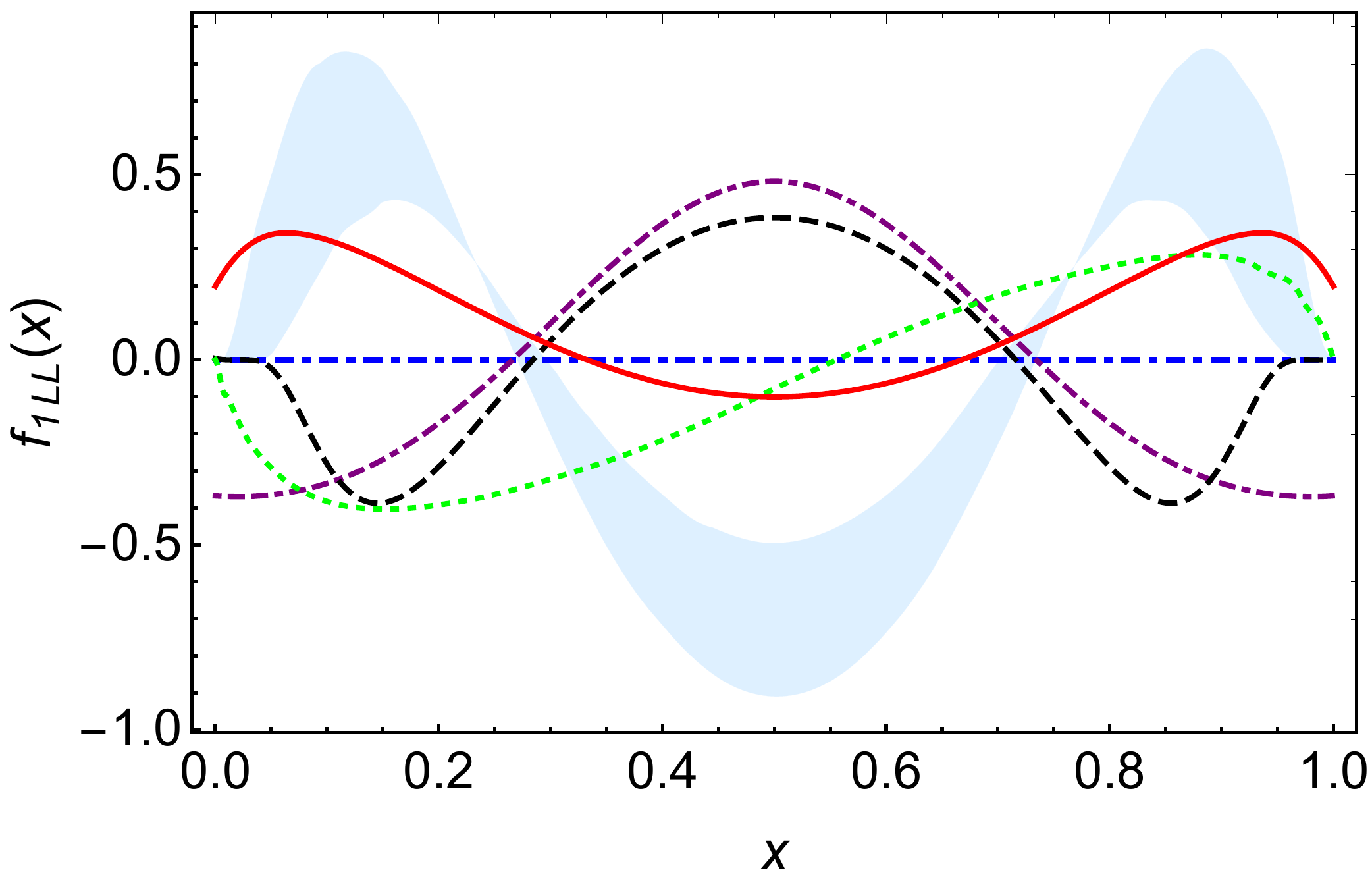} 
\caption{\looseness=-1
The tensor polorized PDF $f_{1LL}(x)$ of $\rho$ from LF constituent quark model \cite{Sun:2017gtz} (green dotted), LF quark model \cite{Kaur:2020emh} (blue dot-dashed), NJL model \cite{Ninomiya:2017ggn}(purple dot-dash-dashed), LF holographic model \cite{Kaur:2020emh}(black dashed), \cite{Mankiewicz:1988dk} (light blue band) and un-rescaled BSE-based LF-LFWFs (red solid). Note that only the large-$x$ part of red solid curve is meaningful. 
}
\label{fig:9}
\end{figure}

Ref. \cite{Best:1997qp} also predicted the moments of valence tensor polarized  PDF $f_{1LL,v}(x)$. However, due to the instability of $f_{1LL}(x)$ under rescaling procedure, we refrain from making prediction on its moments, as the later rely on $f_{1LL}(x)$'s global behavior at all $x$. On the other hand, model studies have given diverse predictions on  $f_{1LL}(x)$ so far, as displayed in Fig.~\ref{fig:9}. In this respect, our LF-LFWFs can shed some light on the $f_{1LL}(x)$'s behavior at relatively large $x$, e.g., $x\gtrsim 0.8$, since the large $x$ behavior of PDFs, i.e., $x \rightarrow 1$, is dominated by LF-LFWFs \cite{Brodsky:2005wx}. So we plot $f_{1LL}(x)$ calculated with our original (un-rescaled) $\rho$ LF-LFWFs as the solid curve in Fig.~\ref{fig:9}. We emphasize that only the large $x$ region of the solid curve is meaningful in such scheme. We find our $f_{1LL}(x)$ is positive at large $x$, similar to results of \cite{Mankiewicz:1988dk,Sun:2017gtz}. Meanwhile, the LF quark model gives vanishing result \cite{Kaur:2020emh}, and the NJL model and LF holographic  model gives negative result \cite{Ninomiya:2017ggn,Kaur:2020emh}. Such discrepancy deserves careful investigation in the future. For instance, a potential solution could be to carry out a fully covariant calculation of $\rho$'s $f_{1LL}$ in analogy to the pion PDF calculation \cite{Bednar:2018mtf}, so that higher Fock-state effects can be counted in.

\section{Summary\label{sec:sum}}

We extend the study of $\rho$ and $J/\psi$ LF-LFWFs in \cite{Shi:2021taf} to extract the  LF-LFWFs of $\Upsilon$ from its BS wave functions. The leading Fock-state approximation is then enforced by rescaling the LF-LFWFs, and the TMDs of $\rho$, $J/\psi$ and $\Upsilon$ are studied using the light front overlap representation. Among all the nine TMDs, the $f_{1LL}$ is unstable under the rescaling  procedure, indicating its sensitivity to Fock-state truncation, thus left out in this study.

For $J/\psi$ and $\Upsilon$, we find the unpolarized, longitudinally polarized and transversely polarized TMDs are all sizable in magnitude, while the tensor polarized TMDs $f_{1LT}$ and $f_{1TT}$ are suppressed. The tensor polarized TMDs change significantly from $J/\psi$ to $\Upsilon$, indicating their sensitivity to the quark mass. All TMDs  in heavy mesons appear symmetric with respect to $x=1/2$ due to the suppressed p- and d-wave LF-LFWFs. 

The $\rho$ TMDs are then explored and compared with existing studies by LF holographic  model, NJL model and LF quark model \cite{Ninomiya:2017ggn,Kaur:2020emh}. The $\rho$ TMDs are generally broad in $x$ and concentrate in low $\vect{k}_T^2$. We point out both LF holographic  model and NJL model have $\Phi^{\Lambda=0}_{|l_z|=1}=\Phi^{\Lambda=\pm 1}_{|l_z|=2}=0$, and after enforcing this condition on the BSE-based LF-LFWFs similar TMDs can be obtained. Our final result on $\rho$ TMDs is given in Fig.~\ref{fig:5}. 
In this case, the  $g_{1T}$ and $h_{1T}$ are found to agree with LF quark model, while the profiles of $f_{1LT}$ and $f_{1TT}$ are new in literature. We therefore argue that LF-LFWFs with higher OAM can have sizable impact in determining certain polarized TMDs. 

Finally, the collinear PDFs of  $\rho$, $J/\psi$ and $\Upsilon$ are studied. We evolve our $\rho$ valence PDFs, i.e., $f_{1,v}(x)$ and $g_{1L,v}(x)$, to the scale of 2.4 GeV. The first three moments of these PDFs are found to be in agreement with lattice prediction within uncertainties \cite{Best:1997qp}. We also calculate the $f_{1LL}(x)$ using un-rescaled BSE-based LF-LFWFs, and find it to be positive at large $x$. A covariant calculation that counts in the higher Fock-states effects on $f_{1LL}(x)$ is thus called for.

\begin{acknowledgments}
We thank Prof. Fan Wang for helpful suggestions. This work is supported by the National Natural Science Foundation of China (under Grant No. 11905104) and the Strategic Priority Research Program of Chinese Academy of Sciences (Grant NO. XDB34030301).
\vspace{3em}
\end{acknowledgments}

\appendix

\section{LF-LFWFs of vector mesons}

\onecolumngrid

\begin{figure}[h!]
\centering\includegraphics[width=0.85\columnwidth]{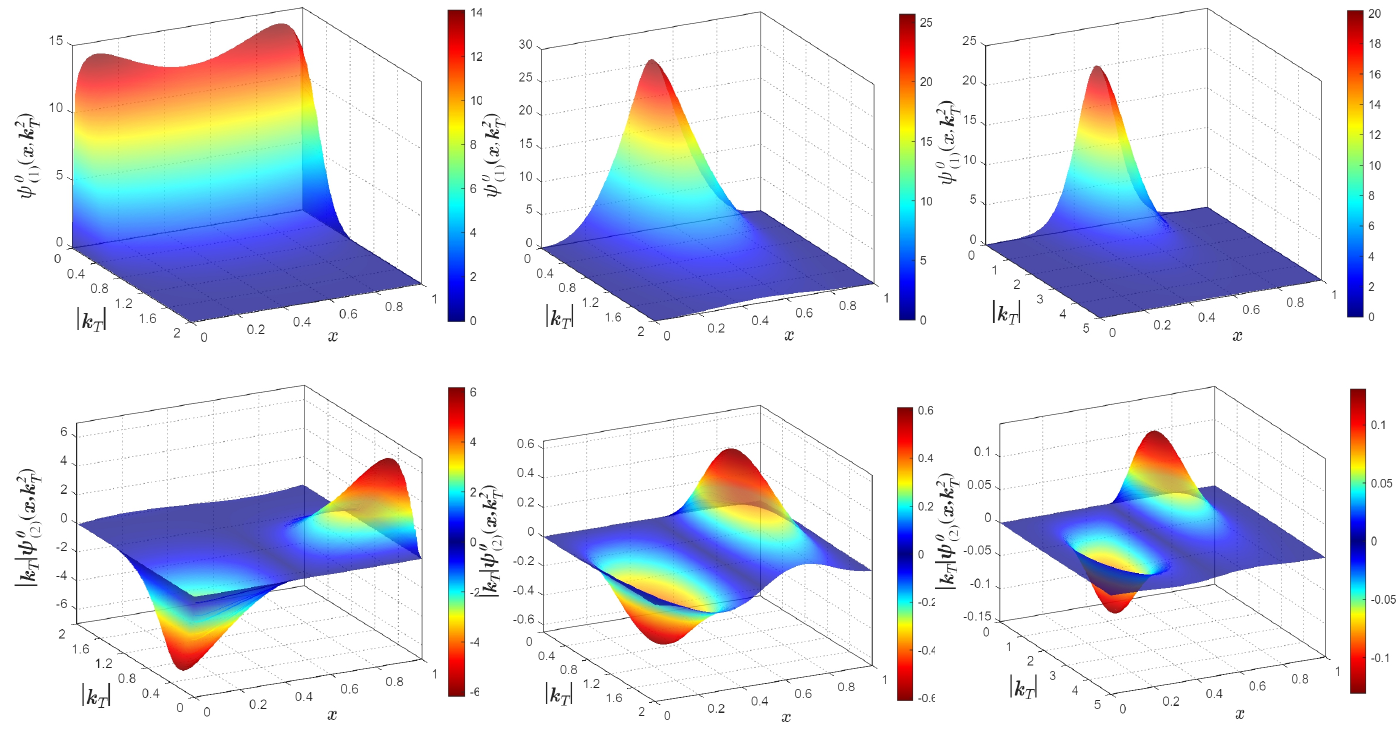} 
\caption{\looseness=-1
The LF-LFWFs of $\rho$ (left column), $J/\psi$ (middle column) and $\Upsilon$ (right column) with $\Lambda=0$. See Eqs.~(\ref{eq:LFWF1},\ref{eq:phi1},\ref{eq:phi2}) for LF-LFWFs' definition. 
}
\label{fig:10}
\end{figure}

\begin{figure}[h!]
\centering\includegraphics[width=0.85\columnwidth]{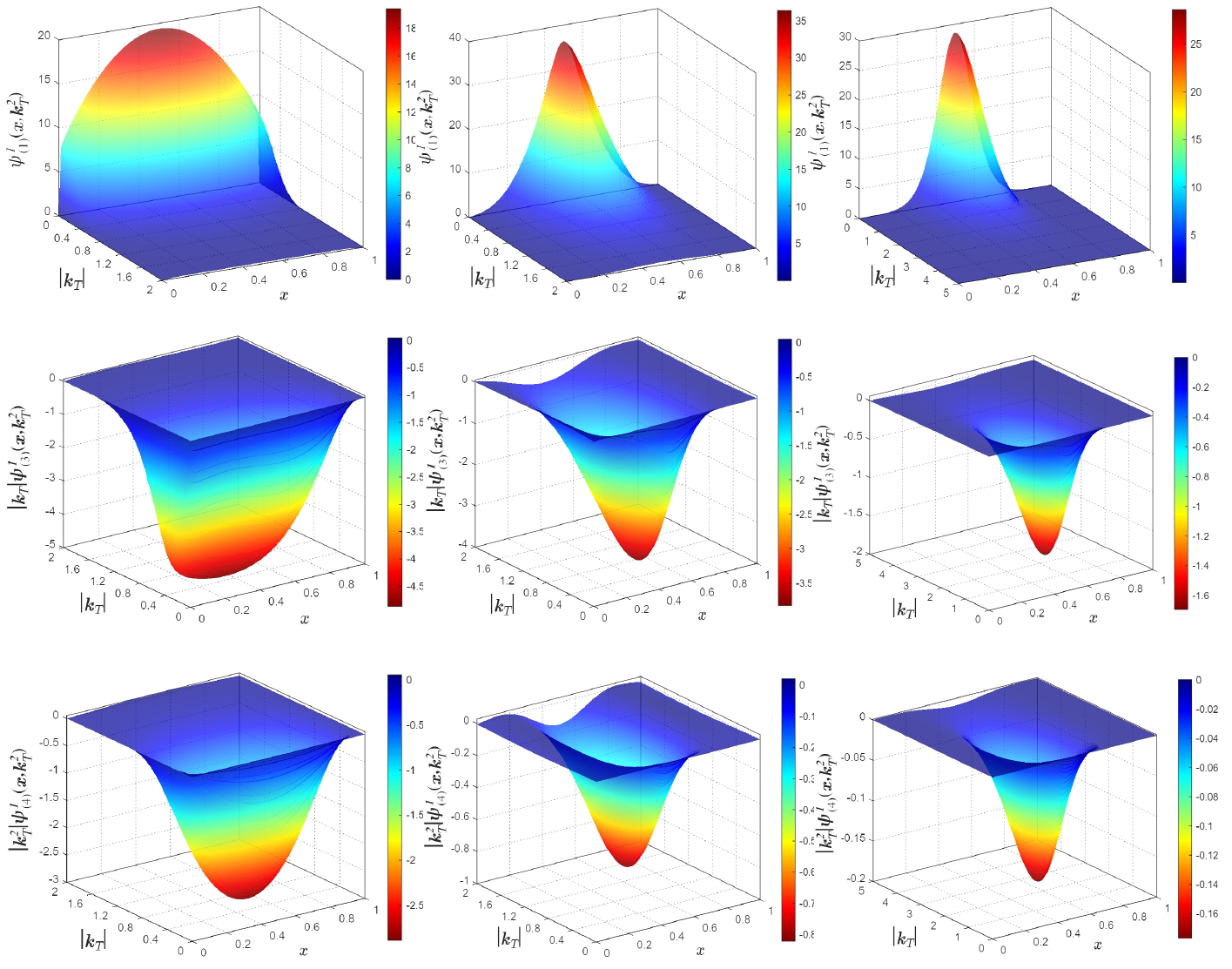} 
\caption{\looseness=-1
The LF-LFWFs of $\rho$ (left column), $J/\psi$ (middle column) and $\Upsilon$ (right column) with $\Lambda=\pm 1$. See Eqs.~(\ref{eq:LFWF1},\ref{eq:phi1},\ref{eq:phi2})  for LF-LFWFs' definition. 
}
\label{fig:11}
\end{figure}

\twocolumngrid

\bibliography{ VTMD}

\end{document}